# Summary of the E&F White Conference
## on
## Radio Frequency Interference Mitigation Strategies

Sydney, Australia, December 1999

Jon F. Bell, Ron D. Ekers

CSIRO ATNF, PO Box 76 Epping NSW 1710 AUSTRALIA;

jbell@atnf.csiro.au, rekers@atnf.csiro.au

John D. Bunton

CSIRO TIP, PO Box 76 Epping NSW 1710, AUSTRALIA;

John.Bunton@tip.csiro.au

10 May 2000

### Abstract

The conference brought together expertise on a range of interference mitigation techniques from CSIRO, Australian and international industry and universities. Key goals were to: enhance the understanding of techniques and their inter-relationship, increase awareness of advanced technologies such as software radios and photonics, and foster a cooperative approach to the development of interference mitigation techniques. The foremost application in mind was the square kilometre array (SKA) and the need to find ways to develop an hierarchical scheme for removing unwanted signals from astronomical data. This paper gives an overview of the topics discussed at the conference and summarises some of the key ideas and results that were presented.

### Introduction

The conference[1] was sponsored by: The Australian Academy of Science[2], CSIRO Australia Telescope National Facility[3], CSIRO Telecommunications and Industrial Physics[4]. In the two days preceding the conference fred harris from San Diego gave a course on digital signal processing[5]. The proceedings of the conference and contact details for all the speakers are available on http://www.atnf.csiro.au/SKA/intmit/atnf/conf/. Further information about interference mitigation is available on http://www.atnf.csiro.au/SKA/intmit/.

---

[1] http://www.atnf.csiro.au/SKA/intmit/atnf/conf/
[2] http://www.science.org.au/
[3] http://www.atnf.csiro.au/
[4] http://www.tip.csiro.au/
[5] http://www.atnf.csiro.au/SKA/intmit/atnf/conf/papers/fharris_dsp_summary.htm

Bob Frater opened the meeting, describing his friendship with Sir Fredrick White whose bequest had made this conference possible. Fred would have been pleased to have this conference under his auspices due to his early close association with the CSIRO Divisions on the Radiophysics site, radio astronomy and his later interest in signal processing.

Ron Ekers then spoke on the future of radio astronomy: options for dealing with human generated interference[1]. Ron highlighted the point that astronomers cannot and do not want to impede the communications revolution, in fact they depend on it to make future experiments possible and affordable. The traditional means of enhancing the sensitivity of radio telescopes (improving bandwidth and system temperature) are near the end of their roads and new ways (including cheap large collecting areas and multiple beam systems) must be sought. Existing spectrum regulations alone will not be sufficient for the interesting radio astronomy questions in the future and methods of dealing with undesired signals have to be implemented. The reason why radio astronomers attempts to do this to date have been rudimentary is that observations have been more seriously limited by factors other than interference. In the future it will no longer be sufficient to simply ignore or bypass interference; radio astronomy must actively combat the problem if future high sensitivity telescopes are to be viable.

The keynote address on radio frequency interference mitigation techniques[1] was given by fred harris. Emphasizing the point that one person's trash is another person's treasure (one person's interference is another person's signal) fred summarised state of the art techniques used in modern communication systems, such as polyphase fft based processors. He then described the key elements of signal processing that can be used in suppressing interference:

- temporal filtering
- spatial filtering
- frequency-azimuth processing
- notching and hole poking
- canceling algorithms

The problems arising from boundary conditions were also covered. Fred concluded by noting that it was important to remember that there is no silver bullet, to watch out for toxic algorithms, and seek the most linear high-speed analog to digital converters.

**Common Ground, Overlapping Interests**

Steve Ellingson kicked off the session, with a presentation on interference mitigation techniques[1]. Steve expanded on the key concepts given by fred, and added the concepts of:
- space-time duality: for every time/frequency domain technique there is a dual space/angle domain technique
- constrained power minimisation (CPM), minimum mean square error (MMSE) and their pitfalls
- parametric techniques and the advantages these can give.

After comparing and contrasting these techniques, Steve made an assessment of the broad range of research fields which need to deal with interference, noting where there are similarities to and differences from radio astronomy. This gave a useful insight into which techniques will and won't be appropriate and Steve summarised some of the most promising prospects. A key point that was made by Steve and others is that in most cases radio astronomers only need to measure the statistical properties of the signal and not the signal itself.

Telecommunications and Radio Astronomy: Synergies and Differences[1] was the subject of a presentation by Alan Young . Alan summarised the directions in which communications systems were developing with increasing frequency, bandwidth and complexity. Basically anything that can be done to minimise redundancy and maximise the use of the spectrum was being hotly pursued. Consequently these signals will be less localised in parameter space and it may be harder to find the orthogonality required to remove them if they are unwanted signals. The most glaring difference between radio astronomy and telecommunications is the amount of money being invested, implying that if radio astronomy is to succeed, it needs to take advantage of cheap components being developed by the communications industry. The fact that there are more low power transmitters means that in general interfering signals will be very close to the telescope. This may open up the possibility of using the wave front curvature as a discriminant. The greater decorrelation across a large array will also help.

Doug Gray gave us an interesting insight into the kinds of techniques used in sonar and radar arrays[1]. A key point Doug emphasised was that he felt there was a much greater synergy between these areas and radio astronomy, than between radio astronomy and telecommunications. Reasons for this include; greater use of arrays (including sparse arrays); more focus on calibration; and the need to deal with non-stationary interferers. Doug attempted to provide a bridge between some of the radio astronomy terminology and that used in sonar and radar systems and emphasised the greater success of adaptive algorithms when they are used in conjunction with stochastic constraints. A few other important synergies and differences came out of the following discussions. Radio astronomers only want the power spectrum in most cases. Most adaptive cancellers provide a reference which is most like the desired signal, while for radio astronomy, the desired signal is unknown, so we must try to get around the problem by providing a reference signal which is most like the interference. Astronomers generally use correlation lags because of the greater physical insight, while radar folks use covariance matrices because they yield insight into neat mathematical tools. Dynamic range was a recurring theme throughout many discussions during the conference. There is a trade off between dynamic range and computational cost.

**Advanced Technologies: Overview and Future Projections**

Peter Hall introduction this session by giving an overview of the relevance to the SKA[1]. Peter emphasised the need for the SKA to use spectral bands outside the traditional radio astronomy bands, and the need for a hierarchical approach for dealing with interfering signals which is:

- effective
- robust
- versatile
- non-toxic to desired signals

Peter also summarised the interference mitigation activities presently underway at ATNF.

The SKA design options[1] presently being considered were presented by Ron Ekers:

| Country | Technique/Technology |
|---|---|
| Netherlands[6] | Planar phased array tiles |
| Australia[7] | Luneburg lenses |
| USA[8] | Commercial Satellite dishes |
| Canada[9] | Large adaptive reflectors |
| China[10] | Arecibo style |

---

[6] http://www.nfra.nl/skai/

[7] http://www.atnf.csiro.au/SKA/

[8] http://www.seti.org/science/1ht.html

[9] http://www.ras.ucalgary.ca/SKA

| India | GMRT style with cheap midsize dishes |

[Fred harris](#) presented the state of the art of [digital receivers](#)[1]. Any real channel has gain, amplitude and phase distortion, frequency dependant fading, Doppler effects, frequency offsets, noise and interfering signals. Digital signal processing provides some novel ways of dealing with these challenges that are not possible in analog devices. A resounding message from fred was that building a digital receiver is not simply a matter of building digital versions of analog components. For example, a conventional analog receiver might apply the operations in the following order:

- frequency shift
- analog low pass filter
- convert to digital

By contrast a well-designed digital receiver that minimises the computational work required might implement things in the following order:

- analog low pass filter
- convert to digital
- resample
- digital low pass filter
- frequency shift

Other points that fred emphasised were: simple is better, real-estate counts, FIR (Finite Impulse Response) filters (nearly) always outperform IIR (Infinite Impulse Response) filters and that multi-rate (input rate not equal to output rate) filters have a lot of advantages. As an example of the state of the art, fred quoted the specs for a Broadcom chip that has a digitiser and digital receiver on a single chip, recording 8-10 bits at 200MHz for a cost of $50 !

Next we received a fascinating insight into [alternative DSP technologies](#)[1] from the past, present and future by Jon Ables. Jon began with stories of the development of the Parkes correlator and Radhakrishnan's famous challenge to think big and build something 10 times bigger than was thought to be possible. (This is something worth bearing in mind as we tackle the R&D challenges of the SKA). Tricks using MLSRS (Maximal Length Shift Register Sequences) to build correlators that are easily scalable in band width and resolution were described along with novel ways of dealing with propagation delays.

[Robert Minasian](#) presented [photonics in radio astronomy: connectivity with in-built signal processing](#)[1], covering topics including:

- Tunable delay lines with delays ranging from 1 ps to 50 ns.
- Dynamic ranges of optical components ranging up to 120dB
- Linearisation
- Tunable notch and band pass filters with 50 - 60 dB rejection and Q factors of 300 - 1000
- FIR filters
- Beam forming arrays
- Future WDM (Wavelength Division Multiplexing) prospects

Design strategies for [high dynamic range receivers](#)[1] were discussed by Russell Gough and George Graves. They summarised the present state of the art radio astronomy systems on the ATNF compact array and noted that they cope with existing levels of interference. The key aspects of good receiver design were covered, as well as future prospects such as high temperature super conducting filters.

---

[10] http://www.bao.ac.cn/bao/LT

[Steve Ellingson](...) presented [a subpace-tracking approach to null forming for large arrays][1] that is being trialed with the OSMA (One Square Metre Array) planar phased array prototype at NFRA in Holland. A key point emphasised by Steve was that beam forming methods like minimum variance (MV) are not appropriate for radio astronomy, because they rely on high INR (interference to noise ratio). Subspace tracking spatial projections (STSP) rely on the ratio of the INR and the SNR, and do not require a reference antenna. There are a number of others features that make these methods preferable to MV, including much more rapid identification of unwanted signals, control of the algorithm when no interference is present, more control over the shape of the main beam. These algorithms are computationally efficient and some FFT based implementations are being developed. They should not be seen as a silver bullet, and there are still some issues to be solved.

**Interference Mitigation Experiences to Date**

[Bob Sault](...) gave an introduction into how [synthesis arrays][1] are used in radio astronomy. Under the assumption that the correlator behaves as an ideal device, the rest of the instrumental effects on the signal phases to be calibrated arise in the antenna and receiver systems. As a result, there are many more equations than unknown parameters, so the system is a closed problem or in radio astronomy jargon, is said to obey closure phase. There are many types of stationary interfering signals that also obey closure phase and if those signals are also enclosed within a delay beam, cross correlating between 3 antennas can provide a reference signal that can be used to cancel the interference. Bob showed some examples of experiments demonstrating this.

Ron Ekers put forward the conjecture "that this post correlation approach and the adaptive cancellers discussed earlier are mathematically equivalent under appropriate conditions". This conjecture is addressed further in presentations by [Steve Ellingson][1] and [Mike Kesteven][1]. Some experiments have been proposed to test the accuracy of this conjecture. [Lisa Kewley](...) then described how this same phase closure technique could be applied [to excising interference from a focal plane array on the Parkes telescope][1]. Lisa showed results of some tests that demonstrated the success of this approach.

A key theme throughout the meeting was that if you have any knowledge of the interfering signal, you could attain much better rejection if you take full advantage of that. At present for radio astronomers, identifying the unwanted signals and determining their characteristics is fairly challenging. [John Sarkissian](...) presented an update on some [transmitter database visualisation][1] software that he has been developing. The aim of this software is to allow astronomers to identify satellite or Australian terrestrial based transmitters that may be sources of interference. In future this database may be closely linked to telescope operation and may be used for automatic detection of interfering signals in the data. Ultimately one could envisage an interfering signal being identified in the data, its modulation and other characteristics being automatically obtained from such a database and then used to excise the unwanted signal.

[David Barnes](...) gave a summary of how the Parkes multi beam receiver system is being used for HI surveys and demonstrated the vast improvements in image quality when [robust statistics][1] are used in the processing of the data. Even a very simple robust statistic such as the median provided far superior data than the mean or sum. Other robust statistics could also be used, but they are more computationally expensive. There are a few disadvantages, including the nonlinear nature of the estimator, higher standard error and the fact that some instrumental or software errors are harder to detect since they are also (partially) rejected.

[Jon Bell](...) described the challenges involved in finding unknown periodic signals from the cosmos and [excision of interfering periodic signals from astronomy data][1]. The dispersed nature of the desired signals provides a natural filter against the unwanted signals, which are not dispersed in general. However the low Q of this filter means that other techniques do need to be employed. The multi beam system on

Parkes used together with a careful observing strategy can provide a further parameter space in which desired and undesired signal are orthogonal in general, allowing the interference to be removed.

It would be useful to test some of the algorithms being proposed on real data. In order to avoid expensive hardware prototyping and hoping to achieve more exacting tests than are possible with simulations, a group at ATNF has been recording base band data containing both astronomical and interfering signals. The idea being that a number of algorithms can be trialed to see which algorithms work best and are less toxic to the astronomical data. John Bunton presented results of some first tests of adaptive filtering on this data[1, 11]. This showed that the astronomical signals could be severely contaminated by such algorithms with only moderate interference suppression, when they are used in an unconstrained way. Using a delayed version of the GLONASS signal seemed to work much better and did not do so much damage to the astronomical signal. There seemed to be a general consensus that constraining such algorithms with a model of the desired or undesired signal would give more satisfactory results. In the weeks following the conference Steve Ellingson applied such a constrained algorithm[12], employing a model of the GLONASS coded chip sequence. This led to considerably improved cancellation with no detectable effects on the astronomical signals.

Mathew Trinkle presented a DSP test bed for interference mitigation in GPS[1] that was under development at CSSIP (Cooperative research centre for Sensor Signal and Information Processing)[13]. The system consists of a 4-element array of omni-directional antennas. The signals are down converted, sampled, passed through the DSP test bed, convert back to analog, up converted and fed into a regular GPS receiver. At present three interference suppression algorithms are implemented in the DSP test bed: and adaptive FIR notch filter, a narrow band beam former and a STAP beam former. The STAP (Space Time Adaptive Processing) beam former allowed a larger number of narrow band interferers to be cancelled with greater rejection.

Reflections on impractical interference mitigation[1] by Jon Ables ranged from the obvious to the ridiculous. However as Jon pointed out in his earlier presentation, crazy suggestions can spur new innovations and make the ridiculous possible. While the far side of the moon is often touted as a suitable place for radio astronomy it has the obvious drawback that one still has to deal with gravity, which inevitably leads to more expensive and more massive construction materials. The Lagrange 2 point, which is a saddle point in the gravitational potential well, formed by the Earth and the sun provides an exciting alternative.

**Application to the SKA**

Lawrence Cram discussed a number of systems considerations[1] in relation to the SKA and interference mitigation, including the basic top level components, such as concentrators, receptors, sub arrays and processors and some of the constraints put on those by the SKA specifications. He raised the problem of whether or not weights from excision or canceling processing in individual stations or sub arrays would need to be recorded so that enough information is available later for the self calibration of the whole array. There was no consensus within the meeting as to whether or not weights need to be recorded. Some simulations or other investigations are needed to sort this out.

Matthew Bailes covered the current state of the art in base band recording systems[1] and possibilities for the future. At the moment a good figure of merit for the affordability of base band recording systems is $3500 per MHz of band width recorded at 2 bits. Commercial off the shelf (COST) approaches (for example DLT (Digital Linear Tape) based recorders) look like providing the best return in the near future and possibly beyond that too. At present we are limited to bandwidths of a few 10's of MHz which make high frequency operation challenging and restricts their use to specialist applications like VLBI and pulsar timing.

---

[11] http://www.atnf.csiro.au/SKA/techdocs/Glonass_cancellation.pdf

[12] http://www.atnf.csiro.au/SKA/intmit/test_data.html

[13] http://www.cssip.edu.au/

The SETI Institute/UC Berkeley rapid prototype array (RPA) [1] is a prototype for the 1hT (1 hectare Telescope), which itself is a prototype for the SKA. Douglas Bock summarised the design specs of the 1hT and then discussed aspects of the RPA, including the backend signal processing, dish mounts, a novel feed design by Jack Welch, and RF signal conditioning. Some substantial site testing and monitoring of interference levels has been undertaken.

John Bunton asked some searching questions in his presentation entitled where do we get the MIPS[14]?[1] The astronomer's wish list for the SKA includes data rates of Gsamples per second per beam. This data stream has to undergo all the usual signal conditioning as well as having undesired interfering signals removed. Current processors can implement simple digital and adaptive filtering at a rate of ~0.01 Gsamples/sec. For 10 beams and 10 interfering sources, the interference excision alone leads to a requirement of 1000 - 10000 high-end processors per SKA station. Accounting for the speed of Intel and correlator processors doubling every 2 years suggests a more manageable number of 10 - 100 processors per antenna station in the year 2012. How far Moore's law will continue beyond this is unclear? Other options include fast programmable gate arrays (FPGAs), which provide more processing power, but with somewhat less flexibility.

Fred harris then lead us through an interesting discussion of fast algorithms[1]. Most people are familiar with the Cooley and Tukey style FFT that gets things done in NlogN operations. Fred explained that using some clever mapping, partitioning and convolution techniques, the Winograd-Rader style algorithms get down to 2N operations.

Most of the people who did not have backgrounds in radar or sonar were having trouble grasping what some of the experts in that area were talking about. A physical perspective of array signal processing[1] presented by Steve Ellingson left most people feeling that they had at least got to first base. If one thinks of the antenna geometry and other signal parameter space as part of a coordinate system, then one can envisage re-mapping or redefining the axis of that system so that the interfering signals are contained within a minimum number of dimensions. In other words obtaining the simplest possible description. Then the eigenvectors describe the output of the array, the auto correlations lie on the diagonal of the covariance matrix and the cross correlations fill in the rest of the matrix. Steve then discussed some simple examples, to illustrate this approach. Steve then applied this approach to the more concrete example of a variation on the Ekers-Sault post-correlation excision[1]. Mike Kesteven also addressed the Ekers conjecture: Pre- and Post- correlation equivalence[1]. Ron Ekers pointed out that if you concentrate on the mathematical aspects you may be able to exploit interesting mathematical techniques, but if you understand the physics you may find other solutions. So both approaches are important. For example, the phase closure approach gives more physical insight but the matrix approach may provide more sophisticated tools.

**Conclusions**

Many of the astronomers who attended the conference were relatively new to many of the interference mitigation techniques that were discussed. The conference has not only made us aware of some good techniques to try, but also will save us some hard work, by making us aware of techniques which will not be useful. [Doug Gray](#) summarised the meeting, giving his thoughts on some of the [best prospects for further consideration](#). In closing, we would like to emphasize that many of the interference challenges we face can be overcome with technical solutions. There will always be a place for regulatory and legal approaches to spectrum management, but too much reliance has been placed on these in the past. The strategy of avoidance has worked well in the past and will continue to be useful in the future, but is now limited by the growth of space based communications systems, which are visible everywhere on Earth. There is no technical solution which is likely to provide a silver bullet. Rather a hierarchical approach will be required, combining a range of techniques. We have just begun exploring technical solutions and their toxicity to astronomical data. We need to continue on this path, determining which techniques are best and how they interact in order to develop a flexible and powerful system for interference suppression.

---

[14] MIPS is an acronym for million instructions per second